\documentclass[aps,prl,twocolumn,superscriptaddress,showpacs,floatfix,10pt]{revtex4-1}
\usepackage{amsmath}
\usepackage{graphicx}
\usepackage{color}
\usepackage{epstopdf}
\usepackage{natbib}
\usepackage{xfrac}
\usepackage[normalem]{ulem}

\begin{document}
\title{Ultrafast and gigantic spin injection in semiconductors}

\author{M.~Battiato}
\email[]{marco.battiato@ifp.tuwien.ac.at}
\affiliation{Institute of Solid State Physics, Vienna University of Technology,  Vienna, Austria}
\author{K.~Held}
\affiliation{Institute of Solid State Physics, Vienna University of Technology,  Vienna, Austria}

\date{\today}

%
%

\begin{abstract} The injection of spin currents in semiconductors is one of the big challenges of spintronics. Motivated by the ultrafast demagnetisation and spin injection into metals,  we propose an alternative  femtosecond route based on the laser excitation of  superdiffusive spin currents  in a  ferromagnet  such as Ni. Our calculations show that even though only a fraction of the current crosses the Ni-Si interface, 
the laser-induced creation of strong transient electrical fields at a ferromagnet-semiconductor interface allows for the injection of chargeless spin currents with a record spin polarisations of 80\%.
Beyond that they are pulsed  on the time scale of 100 femtoseconds which opens the door for new experiments and ultrafast spintronics.
\end{abstract}

\maketitle

Using the  spin rather than the electron's charge for electronics is not
only a fascinating concept but thanks to  Noble prize-winning giant magnetoresistance \cite{FertGruenberg} 
already a billion dollar industry.  Alongside these  metallic
spintronics for information storage, basic research is presently
laying the foundations of semiconductor spintronics for logic devices \cite{Awschalom07} as well as optical \cite{Endres13}, thermoelectric \cite{Breton11,Jeon14}.  Spin currents in semiconductors have a relative 
long lifetime \cite{Kikkawa99} 
and could provide ways to avoid Joule heating \cite{chumak15}. Hence they
prospectively allow for information processing at a high-bandwidth and low energy consumption \cite{Zutic11}.
But even  the first step for such semiconductor spintronics, the spin injection into a semiconductor, preferably  silicon \cite{Appelbaum11,Jansen12,Sverdlov15}, remains a challenge.

The straight-forward idea of spin-polarising a current through a ferromagnet-silicon interface failed by-and-large; it yields a spin polarisation of 0.1\% only \cite{Schmidt00}. More promising are the injection of hot electrons at high energies \cite{Appelbaum07,Appelbaum11} and in particular engineering the ferromagnet-silicon contact by an additional insulating \cite{rashba00,Jonker07,Dash2009} or Schottky barrier \cite{Jansen12}. Despite these successes
 there is  plenty of room for improvements. Also the
 fundamental science point of view calls for a better understanding and
 alternative ways of injecting spin currents.

In a parallel development, the ultrafast spin 
transport in metals has been demonstrated recently and opened an entirely new research field \cite{Malinowski08,Battiato10,Melnikov11,Battiato12,Vodungbo12,Pfau12,Rudolf12,kampfrathnatnano2013,Eschenlohrnatmater2013,Turgut13,Graves13,SchellekensAPL13,Battiato14,Mauchain14,Moisan14,vonKorffSchmising14,Schellekens14,Choi14,SchellekensPRB14,Locht15} whose dawn has been the discovery of ultrafast demagnetisation in Ni \cite{Beaurepaire96}. For instance it was shown how a fs-laser pulse on a ferromagnetic layer could 
inject a spin current into a non-parallel aligned Co-Pt multilayer \cite{Malinowski08}, into non-magnetic Au \cite{Melnikov11}, or into Fe \cite{Rudolf12}. The theoretical study led to the proposal of superdiffusive spin transport \cite{Battiato10,Battiato12,Battiato14} as one microscopic mechanism  for the ultrafast demagnetisation. A number of experiments confirmed the importance of the superdiffusive spin transport \cite{Vodungbo12,Pfau12,Rudolf12,kampfrathnatnano2013,Eschenlohrnatmater2013,Turgut13,Schellekens14,SchellekensPRB14}
as at least  one of the main drivers for the ultrafast demagnetisation.
In particular its power to predict an ultrafast increase of magnetisation \cite{Rudolf12}
and the aforementioned  build up of a magnetisation in a non-ferromagnetic material \cite{Melnikov11} 
established that  superdiffusive spin transport is actually  at work.
While transporting spin information,
the superdiffusive spin injection into a metal is not
a good starting point for spintronic logic devices. The spin diffusion time and length scales
in metals are just too short. Semiconductor devices are needed instead since spin currents survive here on much longer time scales \cite{Kikkawa99}.

In this paper,   the theory of superdiffusive spin transport, 
which successfully predicted the spin injection into metals, is hence extended to a ferromagnet-semiconductor
interface. 

\begin{figure}[tb]
 \includegraphics[width=0.5\textwidth]{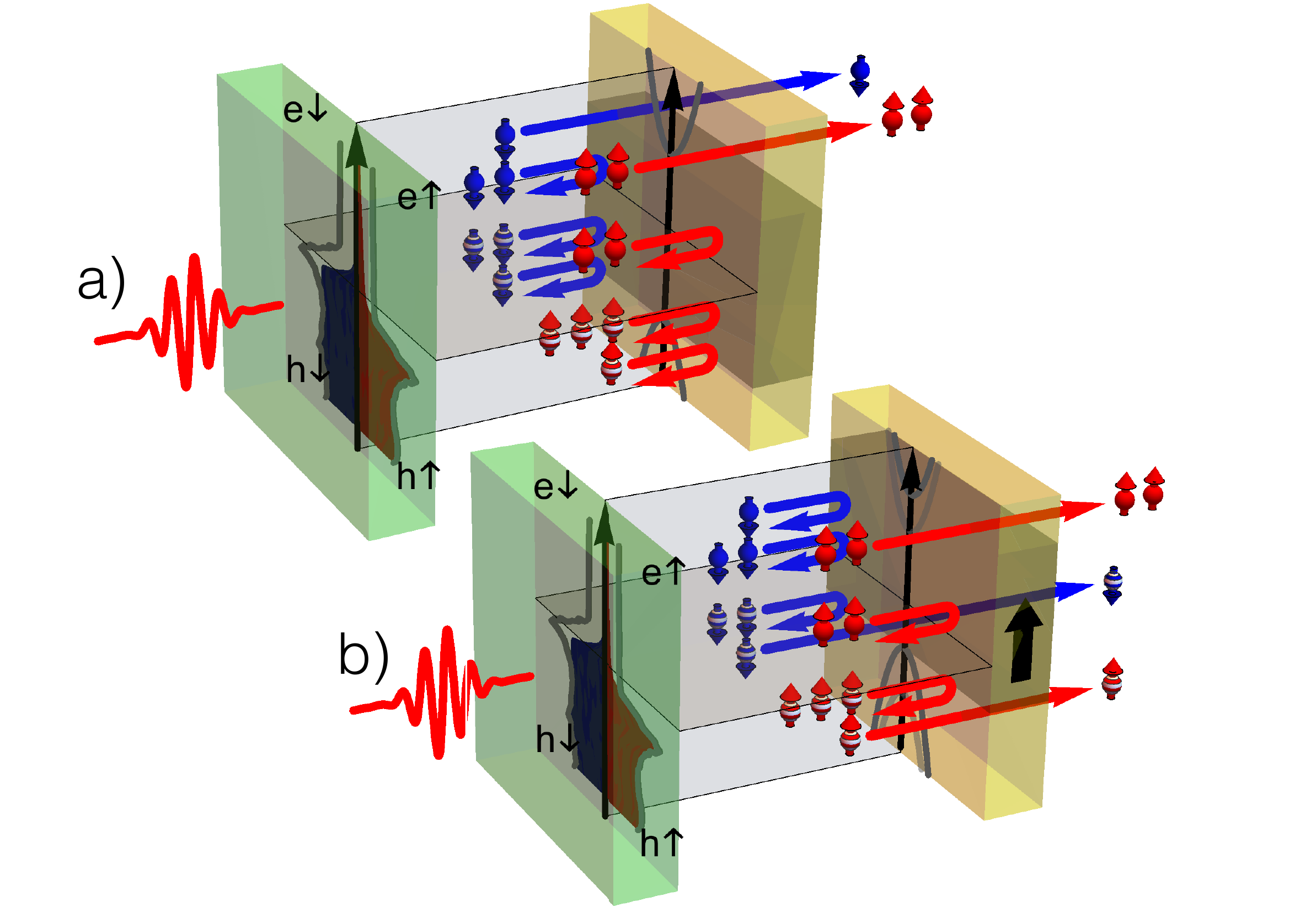}
 \caption{
Sketch of spin injection from ferromagnetic  Ni into a semiconductor  such as Si.
a) Initial situation. The laser pulse excites spin-up (red) and -down (blue) 
electrons (solid spheres)  and holes (dashed spheres); only the 
high energy particles  are not blocked by the Si bandgap (illustrated by two spin-up and one spin-down electrons). b) Almost instantaneously the charge transfer into 
Si shifts the potential upwards until the net charge flow of electrons and holes cancels.
Due to their longer relaxation time there are more high energy spin-up electrons 
so that there is nonetheless a net spin current (indicated by two spin-up electrons and
a spin-down hole vs one spin-down hole entering the semiconductor).
}\label{fig:sketch}
\end{figure}

\textit{General idea.} Let us, before turning to the calculational details, summarise our main findings by hand of Fig.~\ref{fig:sketch}. The  panel a) shows the starting point,
a laser pump pulse is creating a population of excited electrons and holes in Ni.
These charge carriers diffuse through the Ni layer. Of those reaching the Ni-Si interface 
many  are reflected  as indicated in  Fig.~\ref{fig:sketch}  because their energy lies within the Si bandgap (shaded region in Fig.~\ref{fig:sketch}). The electron-hole asymmetry
leads to a large current inflow into  Si. This current is
colossal compared to typical currents  in Si.
Hence it charges the Ni-Si interface. The charge  in turn leads
to a potential step shifting the Si
bandstructure up relative to the Ni one, as envisaged in Fig.~\ref{fig:sketch}b).
The potential up-shift only stops when no further net charge is transferred into Si:
as many electrons as holes enter the Si side.
This process is happening almost instantaneously because of the tremendous current
on the Si surface.

Since Ni is ferromagnetic there is an asymmetry between majority- and minority-spin,  denoted as spin-up and $\mbox{-down}$ in the following. 
Here, the different velocities for spin-up and -down carriers in Ni enter. Even more important is however the different relaxation times. Scattering lifetimes of excited spin-up electrons in Ni are generally much longer than the spin-down electron lifetimes and both, spin-up and -down,  hole lifetimes \cite{Zhukov06}.
Hence, there are much more high-energy spin-up electrons reaching the Ni-Si interface than spin-down electrons as schematically displayed in  Fig.~\ref{fig:sketch}.

After the bandgap shifted  up to the point where as many electrons as holes pass into Si,
on the electron side only the (still unrelaxed) high-energy spin-up electrons are injected as indicated in Fig.\ \ref{fig:sketch}b).
The spin-down electron injection is almost negligible. 
On the hole side there is a smaller spin asymmetry
 so that a similar number of spin-up and -down holes is injected.
This way we have no net charge current into Si, but a sizeable (and pulsed) spin current of $10^{28}  \mu_{\rm B}/({\rm cm}^2 s)$ for an absorbed laser fluence of the order of ${\rm mJ}/{\rm cm}^2$.
This is gigantic in terms of typical charge currents of Si, which should not be surprising since the excitation from a fs laser is a very strong perturbation of the equilibrium and the spin current is present for only a few hundreds of femtoseconds. 

The spin current injected in the Si layer is about two orders of magnitude smaller than the spin current that would be injected in a metallic substrate in a similar configuration \cite{Battiato10,Battiato12,Battiato14}. Nonetheless we can estimate (see more below) that this corresponds to an enormous injection of magnetisation in Si of around $10^{-2}\div 10^{-3} \mu_{\rm B}/{\rm atom}$, which is much larger than a typical concentration of carriers in even heavily doped semiconductors, $10^{-5} {\rm atom}^{-1}$.
Notice that the ultrafast injection of spin in Si from a metal overcomes the problem of conductance mismatch \cite{Schmidt00} as we are in a completely different regime of transport with highly excited charge carriers.   We are now ready to elucidate the two different approximations of the Boltzmann equation used for the metal and the semiconductor, and the strategy for the solution of the coupled problem.

\textit{Method, metal.} In the ferromagnetic metal, 
the femtosecond optical excitation induces  highly excited carriers
whose superdiffusion leads to a spin current due to the asymmetry between the two spin channels. Similarly a charge current is created due to the electron-hole asymmetry. This leads to the build-up of an electric field which, in turn, acts on all the electrons in the metals. 
For convenience we split the carriers into two groups: the highly excited electrons (i.e.~the ones undergoing superdiffusion) and the carriers around the Fermi level (we will refer to them as Fermi carriers). The ample Fermi carriers
dynamically reduce (screen) the electric field.  For our purposes we can  assume, 
as  in Refs.~[\onlinecite{Battiato10}],[\onlinecite{Battiato12}], and [\onlinecite{Battiato14}],  an instantaneous and full compensation of the
electric field, which corresponds to an infinite conductivity of the metal at high frequencies. 
 In this limit the time ($t$) evolution of the carrier distribution $n(z,t,E,\sigma)$ (which also depends on position $z$, energy $E$ and spin $\sigma$)  satisfies the superdiffusive spin transport equation:
\begin{equation} \label{eq:continuity_eq_total_density}
	 \frac{\partial n }{\partial t}= -\frac{n}{\tau}+ \left(-\frac{\partial }{\partial z} \hat{\phi} + \hat{I}\right) \left(\hat{\mathcal{S}}n+S^{ext} \right). 
\end{equation}
The first term on the right hand side describes scattering with lifetimes $\tau(E,\sigma)$. The second term describes the diffusion given by the  superdiffusive flux kernel $ \hat{\phi}$ and the identity operator $\hat{I}$, which act on the laser excited carrier imbalance $S^{ext}(z,t,E,\sigma)$ and on the carrier distribution $n
$ through the scattering operator $\hat{\mathcal{S}}$. For more details see Supplemental Material \cite{supplemental} and Refs.~[\onlinecite{Battiato10}],[\onlinecite{Battiato12}], and [\onlinecite{Battiato14}]


\textit{Method, semiconductor.} The situation is  dramatically different in a semiconductor. For typical carrier injections from the metal the number of excited carriers is (as we will see) much larger than the number of Fermi carriers which can be safely neglected. The superdiffusing carriers will consequently be affected by the bare and non-negligible electrical field.

We are mainly concerned with what happens when a superdiffusive carrier crosses the interface. 
In the case of  metal-metal interfaces  the excited carriers will go through the interface as if it were not present (we neglect scatterings at the interface and band mismatch, see Refs.~[\onlinecite{Battiato10}],[\onlinecite{Battiato12}], and [\onlinecite{Battiato14}]).
When instead an  excited carrier reaches the interface to a semiconductor, two scenarios can be distinguished: a)~If  its energy falls within the bandgap, the carrier will be reflected.  b)~If instead the electron (hole) has an energy above (below) the bandgap of the semiconductor, it will enter the semiconductor conduction (valence) band and continue diffusing according to the transport properties in the semiconductor.  Notice that the bandgap acts in the same way on both spin channels, but its position  leads to an asymmetric reflection of electrons and holes. 

\begin{figure}[t]
 \includegraphics[width=0.49\textwidth]{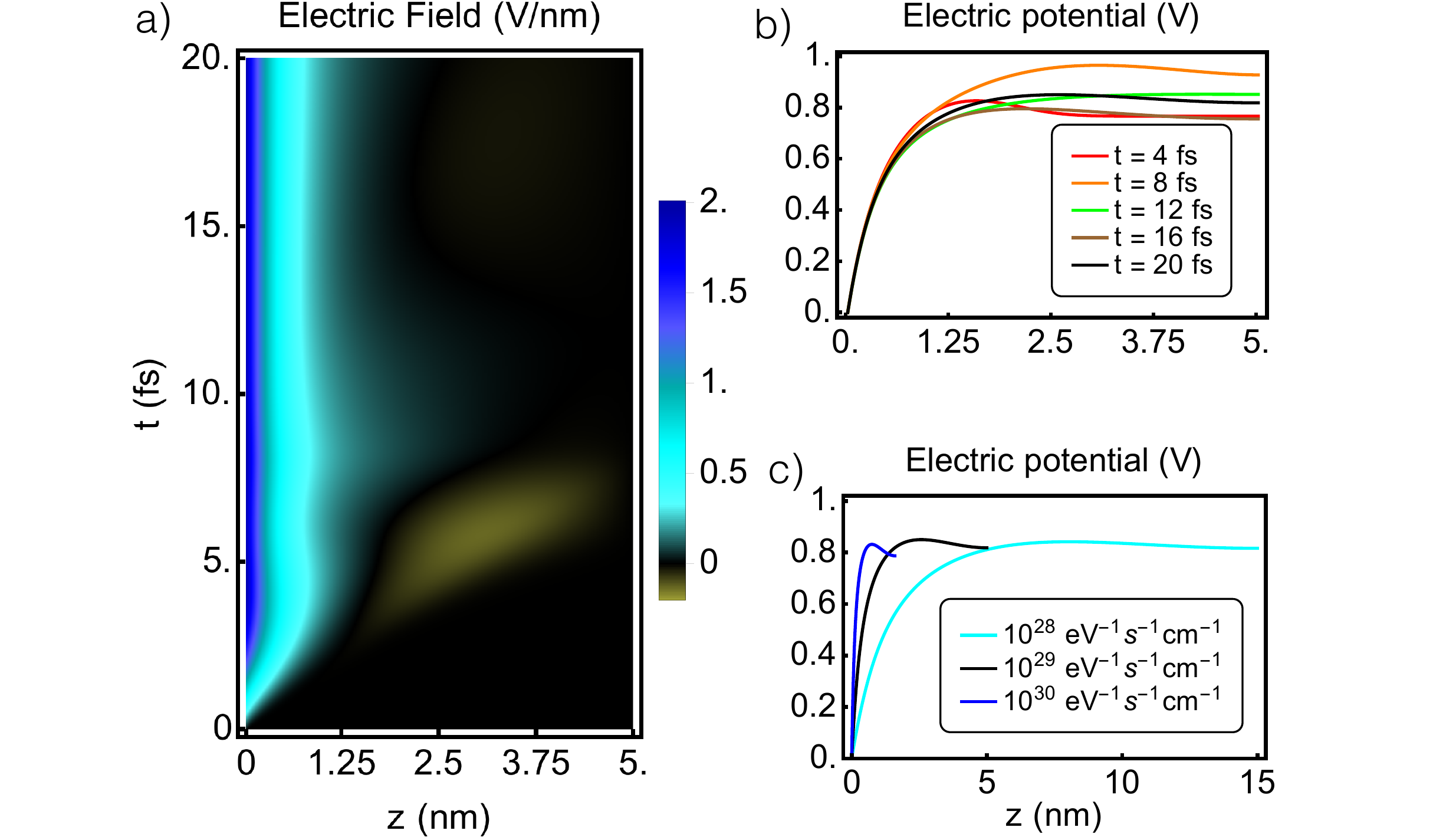}
 \caption{
 Simulation of abruptly switched-on intense net charge injection into a Si layer. a) Position and time resolved amplitude of the electric field. b) Spatial dependence of the electric potential at different times for an electron flux density of $10^{29}$ eV$^{-1}$s$^{-1}$cm$^{-2}$ with electrons (holes) excited up to $1$eV ($0.5$eV). The total hole injection is one order of magnitude smaller than for electrons, see Fig.~\ref{fig:flux}. c) Spatial dependence of the electric potential in the steady state (approximated as the latest simulation time, black line) compared to two further electron fluxes. 
}\label{fig:injSilicon}
\end{figure}

When a carrier manages to cross a metal-semiconductor (M-S) interface, a finite electric charge is removed from the metal and injected into the semiconductor. In the metal, the associated electric field  is immediately screened; the charge hence accumulates at the interface. In the semiconductor instead the charge can spread in space, which in turn will induce a sizeable electric field. This will strongly affect the transport of superdiffusive electrons in Si. 

The coupled solution of the two different types of diffusion in the two parts of the M-S sample is extremely challenging. But a much smarter approach can be followed. We calculate the effect of a strongly asymmetric injection of excited electrons and holes in silicon that is abruptly switched on at time zero. This allows for an estimation of what type of steady state response is reached and of the transient time. In case of short transient time compared to typical dynamics under study, we will be in the position of neglecting the full dynamics of the silicon layer and treating it only as its steady state response. We therefore simulate the injection of an electron flux ten times as high as for holes (compatible with the results we will obtain below in Fig.~\ref{fig:flux}). To this end we employ the Boltzmann equation without scattering but with the electrical field satisfying the Poisson equation, see Supplemental Material \cite{supplemental}. From Fig.~\ref{fig:injSilicon}a) we notice that, at early times, the abundance of injected negative charge charges the Si and causes the appearance of a positive electric field, which acts towards the reduction of the incoming electron flux. A charge shockwave is generated by the abrupt switching on. However the potential profile tends towards the steady state shape already within a few fs and the remaining oscillations decay fast and are hard to distinguish in Fig.~\ref{fig:injSilicon}b) already after $10$fs. Incoming electrons are decelerated by the electric field to negative k's and transported back into the metal outside of the Si; holes are accelerated instead. 

This behaviour depends on the absolute amount of influx in Si. In Fig.~\ref{fig:injSilicon}c) we show calculations for different orders of magnitude of electron flux (the most relevant parameter determining the timescale and the spatial profile). The smaller the flux the slower is the approach to the steady state and the wider the charged region (cf.~Fig.~1 in Supplemental Material \cite{supplemental}). For typical electron fluxes obtained by calculations of superdiffusion in the Ni layer metal (of the order of $10^{29}$ eV$^{-1}$s$^{-1}$cm$^{-2}$) we obtain a transient of a few fs and a charged region of around $1$nm. 

We can therefore safely neglect the transient and suppose that at every instant the Si generates the steady state potential profile instantaneously. An electron entering from the metal will overcome the electric field only if its energy is above the conduction band minimum which is shifted by the electric potential inside the Si. If not, the electron will travel less than $1$nm, invert its direction and be re-injected into the metal after a time of the order of $1$fs. If we neglect the tiny delay, the motion of the electron is identical to a reflection by the bandgap shifted up by the electric potential deep inside the Si. If the charge region is narrow enough the situation is essentially the same for holes, for more details see Supplemental Material \cite{supplemental}. 

A final approximation is based on the fact that lifetimes in the semiconductor are orders of magnitude longer than those in the metal. Since the electrical potential is also nearly flat inside the semiconductor, we can assume that carriers overcoming the electric field within the first few nm in the semiconductor do not return to the metal. Altogether these considerations allow us to describe the M-S interface by superdiffusion in the metal supplemented by  time- and energy-dependent reflection coefficients that correspond to the position of the bandgap in the semiconductor. We are left with the problem of computing the bandgap position. This is however not needed, since we can use the condition for the steady state that no further charge is injected into the semiconductor, i.e.~the net electron flux (injected minus reflected) equals the hole flux.

\textit{Algorithm.} This way the calculation of carrier superdiffusion from time $t$ to time $t+dt$ can be logically split into the following three steps (more in the Supplemental Material \cite{supplemental}): 

(1) compute the evolution from $t$ to $t+dt$ of the superdiffusion in the metallic layer and, in particular, obtain the 
fluxes $\Phi(z_I,t,E,\sigma)$ at the M-S interface ($z=z_I$); 

(2) find the electrical potential $V_0(t)$ (or equivalently the bandgap position) in the semiconductor that corresponds to zero charge injection
\begin{equation} \label{eq:chargflux}
	\sum_{\sigma}\! \Bigg(\!\int\limits_{-\infty}^{E_v(t)} \! \!\! \!\! dE\; \Phi(z_{I},t,E,\sigma) -\!\!\!\int\limits_{E_c(t)}^{+\infty} \!\! \! \!\! dE\; \Phi(z_{I},t,E,\sigma) \Bigg) = 0,
\end{equation}
where $E_v(t)=E_v(0)+ V_0(t)$  ($E_c(t)=E_v(t)+ E_g$) is the top (bottom) of the valence (conduction) band;  and $E_g$ the bandgap;

(3) translate it into energy-dependent reflection coefficients of the M-S interface that are required to compute the superdiffusion in the next timestep (reflection coefficient equal to $1$ for energies within the bandgap and $0$ outside).

 A new and more sophisticated numerical approach had to be developed beyond  the superdiffusive transport of Ref.~[\onlinecite{Battiato10}],[\onlinecite{Battiato12}], and [\onlinecite{Battiato14}], which only included time independent reflection coefficients. 
 
 It is important to notice that while the charge flux in Eq.~(\ref{eq:chargflux}) vanishes, the magnetisation flux
\begin{equation} \label{eq:magnflux}
 \!\Phi_M(t) \! = \!	\sum_{\sigma}\!\sigma \Bigg(\!\int\limits_{-\infty}^{E_v(t)} \!\!\!\! dE \, \Phi(z_{I}\!,\!t\!,\!E\!,\!\sigma) -\!\!\! \!\int\limits_{E_c(t)}^{+\infty} \!\!\!\! dE\,\Phi(z_{I}\!,\!t\!,\!E\!,\!\sigma)\!\! \Bigg) 
\end{equation}
can and actually is finite. 

\textit{Results.} Let us now turn to the results of the superdiffusion calculations in the metal. We consider a layer of Ni with thickness varying from $10$nm to $60$nm deposited on Si. The transport properties of excited carriers in Ni are extracted from \textit{ab initio} calculations \cite{Zhukov06} (see Supplemental Material \cite{supplemental}).  The only relevant parameter required to describe Si is its bandgap,  $E_g=1.1$eV.  

In Fig.~\ref{fig:flux} we report the computed flux through the M-S interface at different energies and times for the Ni[$10$nm]-Si configuration in the case of a $1$eV laser with $2.93$mJ/cm$^2$ fluence. At early times the bottom of the conduction band is very close to the maximum electronic excitation energy.  This is due to the fact that the electrons diffuse faster and reach the  interface in bigger numbers compared to holes. The bandgap  therefore has to move very close to the maximum electron excitation energy to prevent the influx of charge. As  time passes electrons scatter and lower their energy, in particular  high-energy states that have shorter lifetimes than low-energy states. The bandgap  therefore moves towards lower energies. Note that after about 115fs there is essentially no flux of  spin-minority electrons into Si:
due to their shorter lifetime of 2-3 fs (see Supplemental Material \cite{supplemental}) there are no high-energy  spin-minority electrons in Ni soon after the end of the laser pulse at about 100 fs. The majority lifetime is much longer ($>$10 fs). This lifetime-imbalance  is essential for the spin injection.

\begin{figure}[t]
 \includegraphics[width=0.49\textwidth]{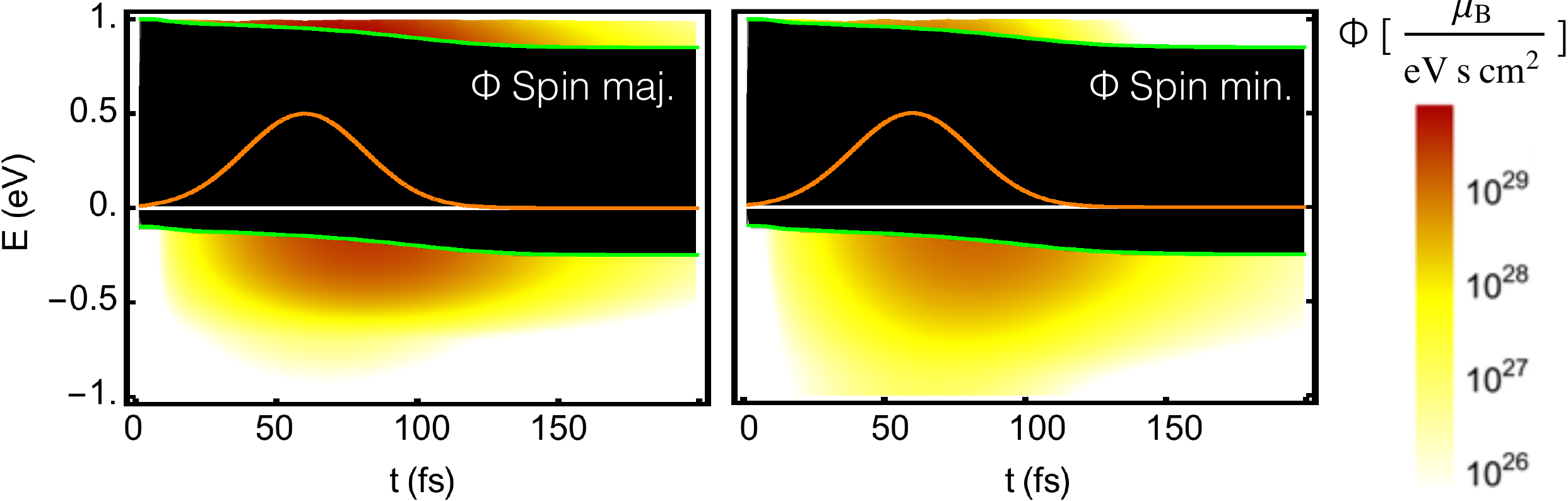}
 \caption{
Energy and time dependent particle flux through the  interface for a Ni[10nm]/Si sample resolved for spin majority (left) and minority (right) carriers. The green lines represent the dynamical position of the top of the valence and bottom of the  conduction band. In the black energy range electrons and holes cannot be injected in the semiconductor. The orange line represents the temporal profile of the pumping laser with  a total fluence of $2.93$mJ/cm$^2$.}\label{fig:flux}
\end{figure}

\begin{figure}[t]
 \includegraphics[width=0.49\textwidth]{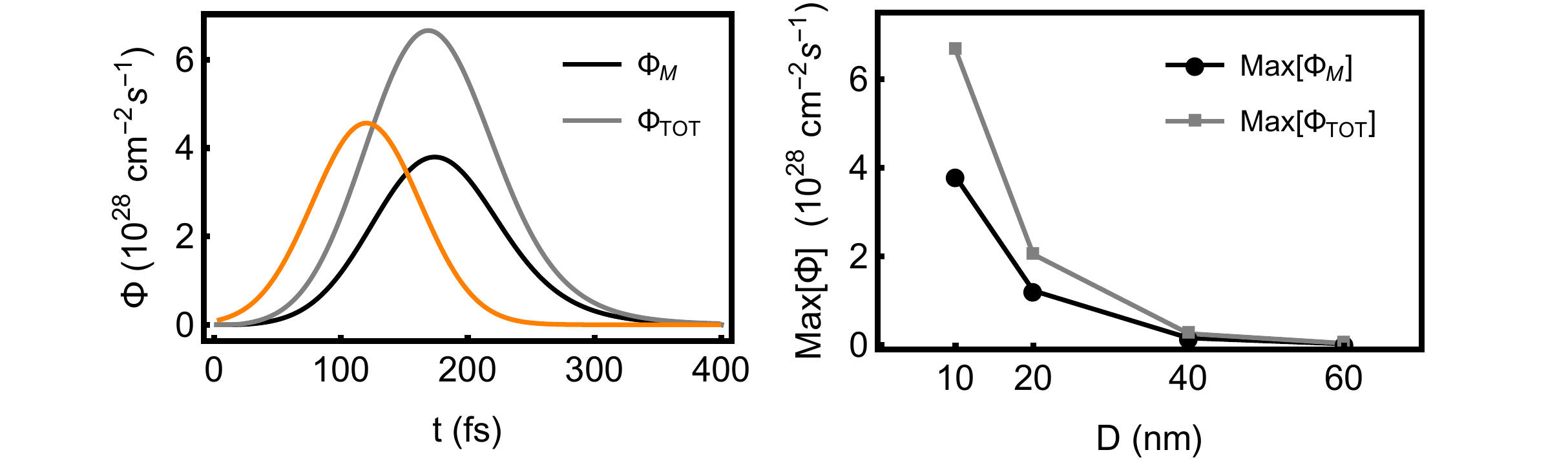}
 \caption{
Left: Magnetisation flux  $\Phi_M$ and total flux $\Phi_{TOT}$ vs.~time for the Ni[10nm]-Si sample and a total laser fluence of $2.93$mJ/cm$^2$. Again the orange line represents the temporal profile of the pumping laser. Right: Maxima of magnetisation and total fluxes vs.~thickness of the Ni film. Thin films are much more suitable for the spin injection.}\label{fig:datathicknesses}
\end{figure}

Fig.~\ref{fig:datathicknesses} (left)
shows the net spin flux $\Phi_M$ through the  interface as a function of time.  The total flux $\Phi_{TOT}$ is around 1.5 times bigger, which implies a spin polarisation of the current of 79\%.  At the beginning the low number of holes reaching the interface acts as a bottleneck to the spin injection. When the holes grow in number due to impact ionisation (cf.\ Fig.~\ref{fig:flux}) their flux can compensate more easily the electron flux which is  predominantly majority-spin. Consequently the spin flux increases dramatically for the 10nm film until $t\sim  180\,$fs. At this point both, electrons and holes, have lost too much energy.
They cannot overcome the bandgap except for  the small high energy tails of the population, which progressively become more and more negligible. The spinflux essentially vanishes at $t=300\,$fs, or $200\,$fs  after the end of the laser excitation.

An important observation is that increasing the Ni thickness dramatically reduces the efficiency of the injection (Fig.~\ref{fig:datathicknesses} right). This is because, during the diffusion towards the M-S interface, the carriers loose some of their energy. Hence, the number of carriers that is able to overcome the bandgap when reaching the M-S interface is strongly reduced for thicker films.

The real striking feature of this kind of injection is the order of magnitude. We obtain that for typical laser fluences used in ultrafast demagnetisation experiments, the magnetic moment injected into the semiconductor is of the order of $10^{-2} \mu_{\rm B}/{\rm atom}$.  This assumes the spin current to be $10^{28}  \mu_{\rm B}/({\rm cm}^2 {\rm s})$ and active for $100 {\rm fs}$. At the same time the carriers will travel approximately in a ballistic way inside the semiconductor at a velocity of the order of $0.5{\rm nm}/{\rm fs}$ and will spread over a length of $0.5{\rm nm}/{\rm fs} \cdot 100 {\rm fs} = 50 {\rm nm}$. The total spin momentum density injected is hence of the order of $10^{28}  \mu_{\rm B}/({\rm cm}^2 {\rm s})\cdot 100 {\rm fs} / 50 {\rm nm}\approx 10^{20}  \mu_{\rm B}/{\rm cm}^3 $. Silicon has approximately $5\cdot 10^{22} {\rm atoms}/ {\rm cm}^3$, leading to a  spin accumulation of around $10^{-2} \mu_{\rm B}/{\rm atom}$ in Fig.\ \ref{fig:datathicknesses} (right) -- which is gigantic for Si.

In conclusion, we have shown how by means of ultrafast optical excitation we can inject a quantity of spin into a semiconductor which is many orders of magnitudes higher than what can be achieved by electrical spin injection. This is because the amount of injected spin is not constrained by the number of charge carriers in the semiconductor but by the amount of excited electrons in the ferromagnetic metal and the spin dependence of its transport properties. 

\begin{acknowledgments}
The authors are grateful to  P.~Maldonado, P.~M.~Oppeneer, M.~M\"unzenberg, and K.~Carva for very fruitful discussions. This work has been supported by the Austrian Science Fund (FWF) through Lise Meitner grant M1925-N28 and SFB ViCoM   F41, as well as by the European Research Council under the European Union's Seventh Framework Program (FP/2007-2013)/ERC through grant agreement n.\ 306447. 
\end{acknowledgments}

\end{document}